\documentstyle[aps,preprint]{revtex}

\begin{document}
\draft
\tighten
\title
{\bf Superconductivity in the Pseudogap State due to Fluctuations of
Short -- Range Order.}
\author{E.Z.Kuchinskii,\ M.V.Sadovskii}
\address
{Institute for Electrophysics,\\ Russian Academy of Sciences, Ural Branch,\\ 
Ekaterinburg,\ 620016, Russia\\
E-mail:\ kuchinsk@iep.uran.ru,\ sadovski@iep.uran.ru} 
\maketitle


\begin{abstract}
We analyze the anomalies of superconducting state ($s$ and $d$--wave pairing)
in a simple model of pseudogap state, induced by fluctuations of short -- 
range order (e.g. antiferromagnetic), based on the model Fermi surface with
``hot patches''. We derive a system of recursion relations for Gorkov's
equations which take into account all diagrams of perturbation theory for
electron interaction with fluctuations of short -- range order. Then we find
superconducting transition temperature and gap behavior for different values 
of the pseudogap width and correlation lengths of short -- range order 
fluctuations.
In a similar approximation we derive the Ginzburg -- Landau expansion and
study the main physical characteristics of a superconductor close to the
transition temperature, both as functions of the pseudogap width and
correlation length of fluctuations. Results obtained are in qualitative 
agreement with a number of experiments on underdoped HTSC -- cuprates.
\end{abstract} 


\newpage
\narrowtext

\section{Introduction.}

The pseudogap state, observed mainly in the underdoped region of the phase
diagram of high -- temperature superconducting (HTSC) cuprates, leads to a
wide range of anomalies of physical properties of these system both in
normal and superconducting states \cite{Tim}. There are two main theoretical
scenarios to explain these anomalies. The first is based upon the model of
Cooper pairs formation already above the temperature of superconducting 
transition \cite{Gesh,EK,Lok} with phase coherence appearing only for $T<T_c$. 
The second assumes that the pseudogap state is induced by fluctuations of
antiferromagnetic (AFM) short -- range order which exist in the underdoped
region of the phase diagram \cite{Kam,Sch,KS}. Recently a number of 
experiments appeared which rather convincingly favor the second scenario
\cite{Lor,Kras}.

Most theoretical papers in the field are dedicated to the study of the models
of the pseudogap state in normal phase for $T>T_c$. 
In Refs. \cite{PS,KSad} we proposed a very simple (toy) exactly solvable 
model of the pseudogap state, which assumes the existence of certain ``hot''
(flat) patches on the Fermi surface. Within this model we derived the 
Ginzburg -- Landau expansion for different types of Cooper pairing \cite{PS}
and studied the anomalies of superconducting state for $T<T_c$ \cite{KSad}
induced by fluctuations of AFM short -- range order. In these papers we 
dealt with (over) simplified model of static Gaussian fluctuations of
short -- range order with infinite correlation length, which allowed us to
obtain an exact analytical solution for the pseudogap state. In real systems
the correlation length of AFM fluctuations is finite and relatively short
\cite{Sch}. The main aim of the current paper is to generalize the main
results of Refs. \cite{PS,KSad} to the case of finite correlation lengths
of fluctuations of AFM short -- range order and to analyze the dependence of
the main physical characteristics of the superconducting state on this
correlation length, as well as on the effective width of the pseudogap.

\section{Model of the Pseudogap State.}

We consider the simplified model of the pseudogap \cite{PS,KSad} based upon
the picture of well -- developed fluctuations of AFM short -- range order and
is in some respects close to the ``hot -- spot'' model of Ref. \cite{Sch}. 
We assume the Fermi surface of two -- dimensional electronic system as shown
in Fig.1. In fact such a Fermi surface was observed in a number of ARPES --
experiments on HTSC -- cuprates \cite{Onell,Shen}. Note that the assumption
of the flatness of these patches is non crucial for our model, but 
significantly simplify calculations, which, in principle, can be done also
in more realistic ``hot-spot'' approach with apparently similar results.
The model Fermi surface as in Fig.1 has already been applied to HTSC --
cuprates in Refs. \cite{Vir,Ruv,Dz}, where the details of microscopic 
criteria for the existence of antiferromagnetic and superconducting phases
were analyzed. Here we just assume a kind of phenomenological model with
static Gaussian fluctuations of short -- range order with correlation 
function (structure factor) of the following form \cite{Kam}:  
\begin{equation} S({\bf q})=
\frac{1}{\pi^2}\frac{\xi^{-1}}{(q_x-Q_x)^2+\xi^{-2}} 
\frac{\xi^{-1}}{(q_y-Q_y)^2+\xi^{-2}}
\label{fluct}
\end{equation}
where $\xi$ -- is correlation length of fluctuations, while the scattering
vector is taken to be $Q_x=\pm 2k_F$,\ $Q_y=0$ or $Q_y=\pm 2k_F$,\ $Q_x=0$, 
anticipating that these fluctuations are incommensurate. The factorized
form of (\ref{fluct}) was first introduced in Ref. \cite{Kam} and leads to
great simplification of all calculations. At the same time we stress that
in practice it practically coincides with the usual isotropic Lorentzian in
the most important region of $|{\bf q-Q}|<\xi^{-1}$ \cite{KS}.

Less justified physically is the assumption of the static nature of
fluctuations which can be reasonable only at high enough temperatures
\cite{Sch,KS}. At low temperatures, including the superconducting phase,
spin dynamics can be very important, e.g. for the microscopics of Cooper
pairing according to the picture of ``nearly -- antiferromagnetic'' Fermi
liquid \cite{MBP,MP}. However, we assume that our static approximation can be
sufficient for qualitative analysis of the influence of superconductivity
upon superconductivity, which will be described below within the standard
BCS -- like approximation.

Let us write down the interaction of electrons with AFM fluctuations in the
following form:
\begin{equation}
V_{eff}=(2\pi)^2W^2S({\bf q})
\label{Veff}
\end{equation}
where $W$ is determining the energy scale (width) of the pseudogap.
We assume that only electrons from the flat (``hot'') of the Fermi surface
interact with AFM fluctuations, so that $W$ is effectively non zero only for
these electrons \cite{PS,KSad}. Note that we completely neglect the spin
structure of interaction which can be rather easily taken into account
\cite{Sch}, but makes calculations more cumbersome. In this sense, strictly
speaking our discussion is more appropriate for the case of electrons
interacting with fluctuations short -- range order of CDW rather than SDW
(AFM) type. We also assume that this simplification is relatively
unimportant for the analysis of qualitative effects of pseudogap on
superconductivity.

The factorized form of the correlator (\ref{fluct}) and of interaction
(\ref{Veff}) leads to the one -- dimensional nature of scattering on 
fluctuations. In the limit of infinite correlation length
$\xi\to\infty$ this model acquires an exact solution \cite{PS,KSad,C74}. 
For finite $\xi$ we can construct ``nearly exact'' solution \cite{KS}, 
directly generalizing the one -- dimensional approach proposed in Ref.
\cite{C79}. In this case we can (approximately) sum the whole diagram
series for one -- particle Green's function of electrons from the flat
parts of the Fermi surface (where the ``nesting'' condition for the
electronic spectrum $\xi_{\bf p\pm Q}=-\xi_{\bf p}$ is satisfied).

For the contribution of an arbitrary diagram for the self -- energy of an
electron of the $N$-th order in (\ref{Veff}) we use the following
{\em Ansatz} \cite{KS,C79}:  
\begin{eqnarray} 
\Sigma^{(N)}(\varepsilon_n{\bf p})=W^{2N}\prod_{j=1}^{2N-1}G_{0k_j}
(\varepsilon{\bf p}), \nonumber\\
G_{0k_j}(\varepsilon_n{\bf p})=\frac{1}{i\varepsilon_n-(-1)^j\xi_{\bf p}+
ik_j\kappa}
\label{Ansatz} 
\end{eqnarray}
where $\kappa=v_F\xi^{-1}$ ($v_F$ -- Fermi velocity), $k_j$ -- is the number
of interaction lines surrounding the $j$-th (from the beginning) electronic
line in diagram, $\varepsilon_n=2\pi T(n+1/2)$ (in the following we write
expressions for the case of $\varepsilon_n>0$).  Thus the contribution of an
arbitrary diagram is determined, in fact, only by some set of integer 
numbers $k_j$. An arbitrary diagram with intersections of interaction lines
is seen to be equal to some diagram without intersections and the 
contribution of all diagrams with intersection can be taken into account
with the help of some combinatorial factors $v(k_j)$, attributed to 
interaction lines on diagrams with no intersections \cite{C79,KS,Sch}. 
In the model of incommensurate fluctuations which we only consider:  
\begin{equation} 
v(k)=\left\{\begin{array}{cc}
\frac{k+1}{2} & \mbox{for odd $k$} \\
\frac{k}{2} & \mbox{for even $k$}
\end{array} \right.
\label{vincomm}
\end{equation}
As a result we obtain the following recursion procedure (continuous
fraction representation) to calculate the one -- particle Green's 
function $G(\varepsilon_n{\bf p})$ of electrons from ``hot patches''
\cite{C79,KS,Sch}:
\begin{equation}
G_k(\varepsilon_n{\bf p})=\frac{1}{i\varepsilon_n-(-1)^k\xi_{\bf p}+
ik\kappa-W^2v(k+1)G_{k+1}(\varepsilon_n{\bf p})};\quad 
G(\varepsilon_n{\bf p})\equiv G_0(\varepsilon_n{\bf p})
\label{Gk}
\end{equation}
Diagrammatically this procedure is shown in Fig.2.

{\em Ansatz} (\ref{Ansatz}) for the contribution of an arbitrary diagram of
the $N$-th order, in general case, is not exact \cite{KS,Tch}. However,
in two -- dimensional case we can explicitly demonstrate the topologies of
the Fermi surface when (\ref{Ansatz}) becomes exact \cite{KS}, for the
general case we can show \cite{KS} that this representation in some sense
overestimates the role of finiteness of correlation length $\xi$ in the
given order of perturbation theory. For one -- dimensional case, when this
problem is particularly serious \cite{KS,Tch}, it appears that for
incommensurate fluctuations calculations of the density of states using
(\ref{Ansatz}) quantitatively almost ideally reproduce \cite{C00} the results
of an exact numerical solution given in Refs. \cite{Kop,Mill}
\footnote{In the case of one -- dimensional problem with commensurate
fluctuations {\em Ansatz} (\ref{Ansatz}) does not describe a certain weak
Dyson -- type singularity of the density of states near the center of the
pseudogap \cite{Kop,Mill}, while outside the region of this singularity it
also gives rather good quantitative description of exact results. Note that
in two -dimensional case  Dyson -- singularity, most probably, is just
absent.}. In the limit of $\xi\to\infty$ {\em Ansatz} (\ref{Ansatz}) reduces
to an exact solution of Ref. \cite{C74}, while in the limit of $\xi\to 0$ 
for the fixed value of $W$ it describes the physically correct limit of free
electrons.

Outside ``hot patches'' electrons in our model do not interact with
fluctuations and the Green's function remains a free one:
\begin{equation}
G(\varepsilon_n{\bf p})=G_{00}(\varepsilon_n{\bf p})=\frac{1}
{i\varepsilon_n-\xi_{\bf p}}
\label{G00}
\end{equation}
This model leads to non Fermi -- liquid like behavior of the spectral
density on ``hot patches'' and to the smeared pseudogap in the 
density of states (cf. similar results in ``hot spots'' model \cite{Sch,KS}).
On ``cold'' parts of the Fermi surface we just get the usual Fermi -- liquid
(free electron) behavior.

\section{Gorkov Equations for a Superconductor with Pseudogap.}

In Refs. \cite{PS,KSad} we have analyzed the anomalies of superconducting
state in an exactly solvable model of the pseudogap state induced by
AFM short -- range order fluctuations with infinite correlation length
($\xi\to\infty$). In particular, in Ref. \cite{KSad} it was shown that
AFM fluctuations may lead to strong fluctuations of superconducting order
parameter (energy gap $\Delta$) breaking the standard assumption of
self -- averaging gap \cite{Gor,Genn,Scloc} which allows independent averaging
(over configurations of the random field of static short -- range order
fluctuations) of the order parameter $\Delta$ and different combinations of
electronic Green's functions, entering the basic equations of the theory.
Usually, the possibility of such independent averaging is supported by the
following argument \cite{Gor,Scloc}: the value of $\Delta$ changes on
characteristic scale of the order of superconducting coherence length
$\xi_0\sim v_F/\Delta_0$ of BCS -- theory, while Green's functions typically
vary on much shorter lengths of the order of interatomic distances.
Naturally, the last assumption becomes wrong if a new length scale 
$\xi\to\infty$ appears in electronic system. However, in case of AFM
correlation length $\xi\ll\xi_0$ (i.e. when AFM fluctuations correlate on
distances much smaller than the size of Cooper pairs) the self -- averaging
property of $\Delta$ remains, being broken only for $\xi>\xi_0$. 
For this reason all our analysis below will be done assuming the 
self -- averaging nature of the energy gap, allowing us to use the standard
approach of the theory of disordered superconductors (mean -- field approach
in terms of Ref. \cite{KSad}). Thus, we temporarily put aside the very
interesting problem of superconductivity in the absence of self --
averageness of the order parameter. Note that in real HTSC -- systems
we, in fact, usually have $\xi\sim\xi_0$, so that these materials are in
the region of parameters most difficult for the theory.

Similarly to Refs. \cite{PS,KSad} we assume that superconducting pairing
is due to the following simplest attraction potential:
\begin{equation}
V({\bf p,p'})=V(\phi,\phi')=-Ve(\phi)e(\phi'),
\label{VV}
\end{equation}
where $\phi$ is the polar angle determining the direction of electronic
momentum ${\bf p}$ in highly conducting plane, while for $e(\phi)$ we take
the simple model dependence:
\begin{equation}
e(\phi)=
\left\{
\begin{array}{ll}
1 & (\mbox{ $s$-wave pairing})\\ 
\sqrt{2}\cos(2\phi) & (\mbox{ $d$-wave pairing})
\end{array}.
\right.
\label{ephi}
\end{equation}
The attraction constant $V$ is, as usual, non zero in some shell of the
width of $2\omega_c$ around the Fermi level ($\omega_c$ -- is characteristic 
frequency of the quanta responsible for electron attraction). In this case
superconducting gap takes the form: 
\begin{equation}
\Delta({\bf p})\equiv \Delta(\phi)=\Delta e(\phi).
\label{DD}
\end{equation}
In the following for brevity we shall denote gap simply as $\Delta$  
instead of $\Delta(\phi)$ writing down explicit angular dependence only
where it is necessary.

In the superconducting state perturbation theory over the interaction with
AFM fluctuations (\ref{fluct}) has to be built on ``free'' normal and
anomalous Green's functions of a superconductor:
\begin{equation}
G_{00}(\varepsilon_n{\bf p})=-\frac{i\varepsilon_n+\xi_{\bf p}}
{\varepsilon_n^2+\xi^2_{\bf p}+|\Delta|^2};\quad
F^+_{00}(\varepsilon_n{\bf p})=\frac{\Delta^*}
{\varepsilon_n^2+\xi^2_{\bf p}+|\Delta|^2}
\label{GoFo}
\end{equation}
In the model with flat patches on the Fermi surface the electronic spectrum
on the patches orthogonal to $p_x$ is:
$\xi_{\bf p}=v_F(|p_x|-p_F)$ as electron velocity ${\bf v}$ is perpendicular 
to $p_y$ (everything is symmetric for patches orthogonal to $p_y$). Thus,
in the case of $s$-wave pairing, when $\Delta$ is independent of the 
direction of electronic momentum, for the model interaction
(\ref{fluct}), (\ref{Veff}) the problem becomes one -- dimensional.
In case of $d$-wave pairing situation is more difficult as even on the
flat patches orthogonal to $p_x$ the value of $\Delta(\phi)$ depends 
on $p_y$ (and symmetrically on patches orthogonal to $p_y$). 
For this reason, in the analysis of $d$-wave pairing it is convenient to
introduce instead of (\ref{fluct}) the fluctuation correlator of the form
\begin{equation} 
S({\bf q})=\frac{1}{\pi} 
\left\{\frac{\xi^{-1}}{(q_x\mp 2p_F)^2+\xi^{-2}}\delta(q_y)+
\frac{\xi^{-1}}{(q_y\mp 2p_F)^2+\xi^{-2}}\delta(q_x)\right\}
\label{fluc}
\end{equation}
In this case interaction does not change $p_y$ or $p_x$ on flat patches
orthogonal to $p_x$ or $p_y$ and the problem again reduces to one -- 
dimensional.

Now we can formulate an analogue of approximation (\ref{Ansatz}) also for
superconducting state. Some details of derivation are given in the Appendix 
A. The contribution of an arbitrary diagram of $N$-th order over interaction
(\ref{Veff}) to normal or anomalous Green's function is given by the product
of  $N+1$ ``free'' normal $G_{0k_j}$ or anomalous $F^+_{0k_j}$ Green's
functions with renormalized frequencies and gaps (cf. below).
Here $k_j$ -- is the number of interaction lines surrounding the 
$j$-th (from the beginning of the diagram) electronic line.
As in normal phase, the contribution of an arbitrary diagram is defined by
the set of integers $k_j$ and each diagram with intersection of interacting
lines is equal to a certain diagram of the same order without intersections.
Thus again we can use only diagrams without intersections of interaction
lines accounting diagrams with intersections by the same combinatorial
factors $v(k)$ (attributed to interaction lines) as in the normal phase.  
As a result we obtain diagrammatic analogue of Gorkov equations
\cite{AGD} shown in Fig.3. Accordingly we get a system of two recursion
relations for normal and anomalous Green's functions:
\begin{eqnarray}
G_k=G_{0k}+G_{0k}\tilde GG_k-G_{0k}\tilde F F^+_k-F_{0k}\tilde G^*F^+_k-
F_{0k}\tilde F^+G_k \nonumber\\
F^+_k=F^+_{0k}+F^+_{0k}\tilde GG_k-F^+_{0k}\tilde FF^+_k+G^*_{0k}\tilde G^*
F^+_k+G^*_{0k}\tilde F^+G_k
\label{Gork}
\end{eqnarray}
where
\begin{equation}
\tilde G=W^2v(k+1)G_{k+1};\quad \tilde F^+=W^2v(k+1)F^+_{k+1}
\label{GF}
\end{equation}
\begin{equation}
G_{0k}(\varepsilon_n{\bf p})=-\frac{i\varepsilon_n+(-1)^k\xi_{\bf p}}
{\tilde\varepsilon_n^2+\xi^2_{\bf p}+|\tilde\Delta|^2};\quad
F^+_{0k}(\varepsilon_n{\bf p})=\frac{\tilde\Delta^*}
{\tilde\varepsilon_n^2+\xi^2_{\bf p}+|\tilde\Delta|^2}
\label{GkFk}
\end{equation}
and we introduced renormalized frequency $\tilde\varepsilon$ and gap 
$\tilde\Delta$ as:
\begin{equation}
\tilde\varepsilon_n=\eta_k\varepsilon_n;\quad \tilde\Delta=\eta_k\Delta;\quad
\eta_k=1+\frac{k\kappa}{\sqrt{\varepsilon_n^2+|\Delta|^2}}
\label{renpar}
\end{equation}
similar to those appearing for superconductors with impurities \cite{AGD}.

From (\ref{Gork})-(\ref{renpar}) it is easy to obtain the system of
recursion relations for real and imaginary parts of normal Green's function,
as well as for anomalous Green's function:
\begin{eqnarray} 
ImG_k=\frac{\tilde\varepsilon-Im\tilde G}{(\tilde\varepsilon-Im\tilde G)^2+
((-1)^k\xi_{\bf p}+Re\tilde G)^2+|\tilde\Delta+\tilde F|^2}\nonumber\\
ReG_k=\frac{(-1)^k\xi_{\bf p}+Re\tilde G}{(\tilde\varepsilon-Im\tilde G)^2+
((-1)^k\xi_{\bf p}+Re\tilde G)^2+|\tilde\Delta+\tilde F|^2}\nonumber\\
F^+_k=\frac{\tilde\Delta^*+\tilde F^+}{(\tilde\varepsilon-Im\tilde G)^2+
((-1)^k\xi_{\bf p}+Re\tilde G)^2+|\tilde\Delta+\tilde F|^2}
\label{imregf}
\end{eqnarray}
Let us introduce the following notations:
\begin{equation}
ImG_k=-\varepsilon_n J_k;\quad
ReG_k=-(-1)^k\xi_{\bf p}R_k;\quad
F^+_k=\Delta^*f_k
\label{JRf}
\end{equation}
Then we can see that recursion relations for $J_k$ and $f_k$ just coincide,
so that $J_k=f_k$. Finally we obtain the following system of recursion
relations for  $J_k$ and $R_k$:
\begin{eqnarray}
J_k=\frac{\eta_k+W^2v(k+1)J_{k+1}}{(\varepsilon_n^2+\Delta^2)
(\eta_k+W^2v(k+1)J_{k+1})^2+\xi_{\bf p}^2(1+W^2v(k+1)R_{k+1})^2}\nonumber\\
R_k=\frac{1+W^2v(k+1)R_{k+1}}{(\varepsilon_n^2+\Delta^2)
(\eta_k+W^2v(k+1)J_{k+1})^2+\xi_{\bf p}^2(1+W^2v(k+1)R_{k+1})^2}
\label{JR}
\end{eqnarray}
Then the normal and anomalous Green's functions of a superconductor are
determined through $R_0$ and $J_0$:
\begin{equation}
ImG=-\varepsilon_nJ_0;\quad
ReG=-\xi_{\bf p}R_0;\quad
F^+=\Delta^*J_0
\label{scgrin}
\end{equation}
and represent the complete sum of perturbation series for electron in a
superconductor interacting with AFM short -- range order fluctuations.

\section{Critical Temperature and Temperature Dependence of the Gap.}

Energy gap of a superconductor is determined by the equation:
\begin{equation}
\Delta({\bf p})=-T\sum_{\bf p'}\sum_{\varepsilon_n}V_{sc}({\bf p,p'})
F(\varepsilon_n{\bf p'})
\label{Gapeq}
\end{equation}
On the flat parts of the Fermi surface the anomalous Green's function is
defined by (\ref{scgrin}) and recursion relations (\ref{JR}). On the rest
(``cold'' part) of the Fermi surface the scattering on AFM fluctuations is
absent (in our model), so that the anomalous Green's there function is given
by (\ref{GoFo}). As a result, for the case of $s$-wave pairing, with the
account of (\ref{ephi}), the gap equation (\ref{Gapeq}) takes the form:
\begin{equation}
1=\lambda\Biggl\{\tilde\alpha 
T\sum_{\varepsilon_n}\int_{-\omega_c}^{\omega_c} d\xi 
J_0(\varepsilon_n\xi)+(1-\tilde\alpha)\int_{0}^{\omega_c}d\xi 
\frac{th\frac{\sqrt{\xi^2+\Delta^2}}{2T}}{\sqrt{\xi^2+\Delta^2}}\Biggr\}
\label{sgap}
\end{equation}
where $\lambda=VN_0(0)$ is dimensionless coupling constant of pairing
interaction ($N_0(0)$ -- free -- electron density of states at the Fermi level 
), $\tilde\alpha=4\alpha/\pi$, where $\alpha$ is the angular size of a flat
patch on the Fermi surface (cf. Fig.1). In numerical calculations below
we, rather arbitrarily, choose $\tilde\alpha=2/3$, i.e.
$\alpha=\pi/6$, which is close e.g. to the data of Ref. \cite{Onell}.

In case of $d$-wave pairing we have to take into account the angular
dependence of the gap (\ref{DD}), so that Eq. (\ref{Gapeq}) becomes:
\begin{equation}
1=\lambda\frac{4}{\pi}\Biggl\{T\int_{0}^{\alpha}d\phi e^2(\phi)
\sum_{\varepsilon_n}\int_{-\omega_c}^{\omega_c} d\xi 
J_0(\varepsilon_n\xi)+\int_{\alpha}^{\pi/4}d\phi e^2(\phi)
\int_{0}^{\omega_c}d\xi\frac{th\frac{\sqrt{\xi^2+\Delta^2e^2(\phi)}}{2T}}
{\sqrt{\xi^2+\Delta^2e^2(\phi)}}\Biggr\}
\label{dgap}
\end{equation}
In Fig.4 we show the calculated (from Eq. (\ref{sgap})) temperature
dependences of the gap for the case of $s$-wave pairing and for different
values of correlation length (parameter $\kappa=v_F\xi^{-1}$) of 
fluctuations. For the case of $d$-wave pairing results are qualitatively
similar.

Equation for superconducting critical temperature $T_c$ immediately follows
from (\ref{sgap}), (\ref{dgap}) as $\Delta\to 0$. In this case
$J_0(\Delta\to 0)$ is independent of  $\phi$ and is the same both for
$s$ and $d$-wave pairing. Then $T_c$ -- equation takes the following form:
\begin{equation}
1=\lambda\Biggl\{\alpha_{eff}T_c\sum_{\varepsilon_n}\int_{-\omega_c}^{\omega_c}
d\xi J_0(\varepsilon_n\xi;\Delta\to 0)+(1-\alpha_{eff})\int_{0}^{\omega_c}
d\xi\frac{th\frac{\xi}{2T_c}}{\xi}\Biggr\}
\label{Tc}
\end{equation}
where an ``effective'' size of flat patches on the Fermi surface is defined
as:
\begin{equation}
\alpha_{eff}=
\left\{
\begin{array}{ll}
\tilde\alpha & (\mbox{ $s$-wave pairing})\\ 
\tilde\alpha+\frac{1}{\pi}\sin (\pi\tilde\alpha ) & (\mbox{ $d$-wave pairing})
\end{array}.
\right.
\label{aleff}
\end{equation}
Calculated dependences of $T_c$ on the width of the pseudogap $W$ and
correlation length (parameter $\kappa=v_F\xi^{-1}$) are shown in Fig.5
($T_{c0}$ -- transition temperature in the absence of pseudogap).

The general qualitative conclusion is the same as in Refs. \cite{PS,KSad}:
pseudogap suppresses superconductivity due to a partial 
``dielectrization'' of electronic spectrum on ``hot'' parts of the Fermi
surface. This suppression effect is maximal for  $\kappa=0$ (infinite 
correlation length of AFM fluctuations) \cite{PS,KSad} and  weakens as
correlation length becomes shorter. These results are in full accordance
with experimental phase diagram of HTSC -- cuprates.

Let us stress once again that all our results are valid in case of
self -- averaging superconducting order parameter (mean -- field approach
of Ref. \cite{KSad}), which is valid for not very large correlation lengths
$\xi<\xi_0$, where $\xi_0$ -- is superconducting coherence length
(the size of Cooper pairs at $T=0$). For $\xi\gg\xi_0$ important effects
due to non self -- averaging gap fluctuations appear, leading e.g to 
characteristic ``tails'' in temperature dependence of the average gap 
for $T_c<T<T_{co}$ \cite{KSad}.

\section{Cooper Instability. Recurrence Procedure for the Vertex Part.}

It is well known that the critical temperature can also be determined from
Cooper instability of the normal phase:
\begin{equation}
1-V\chi(0,0)=0
\label{coopinst}
\end{equation}
where the generalized Cooper susceptibility is defined by the graph shown in
Fig.6. Here we have to calculate the ``triangular'' vertex part accounting
for interaction with AFM fluctuations. For the similar one -- dimensional
problem (and for real frequencies $T=0$) the appropriate recurrence procedure
was formulated in Ref. \cite{C91}. For our two -- dimensional model this
procedure was used in Ref. \cite{C99} to calculate optical conductivity.
Generalization to Matsubara frequencies is rather straightforward. Below, for
definiteness, we assume $\varepsilon_n>0$. Then we obtain:
\begin{eqnarray}
\Gamma_{k-1}(\varepsilon_n,-\varepsilon_n,{\bf q})=1+\nonumber\\
+W^2v(k)G_k\bar G_k
\Biggl\{1+\frac{2ik\kappa}{2i\varepsilon_n-(-1)^kv_Fq-W^2v(k+1)
(G_{k+1}-\bar G_{k+1})}\Biggr\}\Gamma_{k}(\varepsilon_n,-\varepsilon_n,
{\bf q})  
\nonumber\\
\Gamma(\varepsilon_n,-\varepsilon_n,{\bf q})\equiv 
\Gamma_{0}(\varepsilon_n,-\varepsilon_n,{\bf q})
\label{Gamma}
\end{eqnarray}
where  $G_k=G_k(\varepsilon_n{\bf p+q})$ 
and $\bar G_k=G_k(-\varepsilon_n,{\bf p})$ are calculated from (\ref{Gk}).

To find $T_c$ we have to know the vertex part at ${\bf q}=0$. Then
$\bar G_k=G^*_k$ and the vertex $\Gamma_k$ becomes real, which considerably
simplifies (\ref{Gamma}). Using notations similar to (\ref{JRf}), 
from (\ref{Gk}) and (\ref{Gamma}) we get:
\begin{equation}
\Gamma_{k-1}=1+W^2v(k)\frac{J_k}{1+W^2v(k+1)J_{k+1}}\Gamma_{k}
\label{Gammk}
\end{equation}
while for $R_k$ and $J_k$ we have recursion relations given by (\ref{JR}) 
with $\Delta=0$.

There exists the following exact relation similar to the Ward identity 
(the proof of this relation will be given below):
\begin{equation}
G(\varepsilon_n{\bf p})G(-\varepsilon_n{\bf p})\Gamma(\varepsilon_n,
-\varepsilon_n,0)=(\xi^2_{\bf p}R_0^2(\varepsilon_n\xi_{\bf p})+
\varepsilon_n^2J_0^2(\varepsilon_n\xi_{\bf p}))\Gamma_0(\varepsilon_n,
-\varepsilon_n,0)\equiv J_0(\varepsilon_n\xi_{\bf p})=
-\frac{ImG(\varepsilon_n{\bf p})}{\varepsilon_n}
\label{Ward}
\end{equation}
Numerical analysis fully confirms this relation, demonstrating the self --
consistency of our recursion relations for one -- particle Green's function
and vertex part
\footnote{Note that an analytic proof of this relation from direct
comparison of recursion procedures for Green's function and vertex part is
non obvious, to say the least.}. As $J_0(\Delta\to 0)$ coincides with
$J_0$ in normal phase, the relation (\ref{Ward}) leads to $T_c$ -- equation 
obtained from Cooper instability (\ref{coopinst}):  
\begin{equation} 
1=\lambda\Biggl\{\alpha_{eff}T_c\sum_{\varepsilon_n}\int_{-\omega_c}^{\omega_c}
d\xi(\xi^2_{\bf p}R_0^2(\varepsilon_n\xi_{\bf p})+
\varepsilon_n^2J_0^2(\varepsilon_n\xi_{\bf p}))\Gamma_0(\varepsilon_n,
-\varepsilon_n,0)+(1-\alpha_{eff})\int_{0}^{\omega_c}d\xi
\frac{th\frac{\xi}{2T_c}}{\xi}\Biggr\}
\label{Tcc}
\end{equation}
being the same as Eq. (\ref{Tc}), obtained by linearization of the gap
equation, despite seemingly different recursion procedures used to obtain
these equations.

\section{Ginzburg -- Landau Expansion.}

In Ref. \cite{PS} we derived the Ginzburg -- Landau expansion in exactly
solvable model of the pseudogap with infinite correlation length of AFM
fluctuations. Here we generalize these results to the case of finite
correlation lengths.

Let us write the Ginzburg -- Landau expansion for the difference of free
energies of superconducting and normal state in the following form:
\begin{equation}
F_{s}-F_{n}=A|\Delta_{\bf q}|^2
+q^2 C|\Delta_{\bf q}|^2+\frac{B}{2}|\Delta_{\bf q}|^4,
\label{GL}
\end{equation}
where $\Delta_{\bf q}$ -- is the amplitude of the Fourier -- component of the
order parameter
\begin{equation}
\Delta(\phi,{\bf q})=\Delta_qe(\phi).
\label{FF}
\end{equation}
Now (\ref{GL}) is defined by diagrams of loop -- expansion for the free
energy of an electron in the field of fluctuations of the order parameter
with small wave -- vector ${\bf q}$ \cite{PS}. 

We express the coefficients of Ginzburg -- Landau expansion as:
\begin{equation}
A=A_{0}K_{A};\qquad   C=C_{0}K_{C};\qquad    B=B_0K_B,
\label{ACD}
\end{equation}
where $A_{0}$, $C_{0}$ and $B_0$ denote the standard expressions for these 
coefficients in the case of isotropic $s$-wave pairing:
\begin{equation}
A_{0}=N_0(0)\frac{T-T_{c}}{T_{c}};\qquad
C_{0}=N_0(0)\frac{7\zeta(3)}{32\pi^{2}}\frac{v_F^2}{T_c^2};\qquad
B_0=N_0(0)\frac{7\zeta(3)}{8\pi^{2}T_c^2},
\label{ACDf}
\end{equation}
Then all the anomalies of the model under consideration, connected with the
appearance of the pseudogap, are contained in dimensionless coefficients
$K_{A}$, $K_{C}$ and $K_B$.
In the absence of pseudogap all these coefficients are equal to 1, only in
case of $d$-wave pairing we have $K_B=3/2$. Thus for $d$- wave pairing we
shall appropriately normalize $K_B$, giving numerical results for
$\tilde K_B= 2/3K_B$.

Consider the generalized Cooper susceptibility shown in Fig.6.
\begin{equation}
\chi({\bf q}0;T)=-T\sum_{\varepsilon_n}\sum_{\bf p}G(\varepsilon_n{\bf p+q})
G(-\varepsilon_n{\bf p})e^2(\phi)\Gamma(\varepsilon_n,-\varepsilon_n,\bf q)
\label{chiq}
\end{equation}
Using (\ref{Ward}) we can express the coefficients $K_A$ and $K_C$ as:
\begin{eqnarray}
K_A=\frac{\chi({\bf q}0;T)-\chi(00;T_c)}{A_0}=\nonumber\\
=\alpha_{eff}\frac{T_c}{T-T_c}\Biggl\{T\sum_{\varepsilon_n=\pi T(2n+1)}
\int_{-\omega_c}^{\omega_c}d\xi J_0(\varepsilon_n\xi)-
T_c\sum_{\varepsilon=\pi T_c(2n+1)}\int_{-\omega_c}^{\omega_c}d\xi 
J_0(\varepsilon_n\xi)\Biggr\}+1-\alpha_{eff}
\label{Ka}
\end{eqnarray}
\begin{eqnarray}
K_C=\lim_{q\to 0}\frac{\chi({\bf q}0;T_c)-\chi(00;T_c)}{q^2C_0}=\nonumber\\
=\frac{32\pi^2T_c^3}{7\zeta(3)v_Fq^2}
\alpha_{eff}
\Biggl\{\sum_{\varepsilon_n=\pi T_c(2n+1)}
\int_{-\omega_c}^{\omega_c}d\xi J_0(\varepsilon_n\xi)-\nonumber\\
-\sum_{\varepsilon=\pi T_c(2n+1)}\int_{-\omega_c}^{\omega_c}d\xi 
G(\varepsilon_n,\xi+\frac{1}{2}v_Fq)\Gamma(\varepsilon_n,-\varepsilon_n,q)
G(-\varepsilon_n,\xi-\frac{1}{2}v_Fq)\Biggr\}
+1-\alpha_{eff}
\label{Kc}
\end{eqnarray}
Situation with coefficient $B$, in general case, is more complicated. 
Important simplifications appear if we limit ourselves in the order of
$|\Delta_q|^4$, as is usually done, by considering only the case of
$q=0$. Then the coefficient $B$ can be determined directly from the anomalous
Green's function $F$, for which we already have the recurrence procedure
(\ref{JR}), (\ref{scgrin}). Let us consider the diagrammatic expansion of the
anomalous Green's function, shown in Fig.7(a). From this it becomes clear that:
\begin{equation}
\lim_{\Delta\to 0}\frac{F(\varepsilon_n{\bf p})}{\Delta}=
G(\varepsilon_n{\bf p})G(-\varepsilon_n{\bf p})+...=
G(\varepsilon_n{\bf p})G(-\varepsilon_n{\bf p})\Gamma(\varepsilon_n,
-\varepsilon_n,0)
\label{wardpr}
\end{equation}
which, by the way, immediately proves (\ref{Ward}) (use also (\ref{scgrin})).
Thus, for the two -- particle loop $\chi(0,0)$ we get:
\begin{equation}
\chi(0,0)=T\sum_{\bf p}\sum_{\varepsilon_n}\lim_{\Delta\to 0}
\frac{F(\varepsilon_n{\bf p})}{\Delta}=T\sum_{\bf p}\sum_{\varepsilon_n}
J_0(\Delta=0)
\label{chi00}
\end{equation}
For the ``four -- tail'' diagram of Fig.7(b), determining the coefficient
$B$, in the same way we obtain:
\begin{equation}
-T\sum_{\bf p}\sum_{\varepsilon_n}\lim_{\Delta\to 0}
\frac{\frac{F(\varepsilon_n{\bf p})}{\Delta}-\lim_{\Delta\to 0}
\frac{F(\varepsilon_n{\bf p})}{\Delta}}{|\Delta|^2}=
-T\sum_{\bf p}\sum_{\varepsilon_n}\lim_{\Delta\to 0}
\frac{J_0(\Delta)-J_0(\Delta=0)}{|\Delta|^2}
\label{4tail}
\end{equation}
where $J_0(\Delta)$ is defined by the recursion procedure (\ref{JR}).
Finally, for the dimensionless coefficient $K_B$ we have:
\begin{equation}
K_B=\alpha_B\frac{8\pi^2T_c^3}{7\zeta(3)}\sum_{\varepsilon_n}
\int_{-\omega_c}^{\omega_c}d\xi \lim_{\Delta\to 0}
\frac{J_0(\Delta=0)-J_0(\Delta)}{|\Delta|^2}+1-\alpha_B
\label{Kb}
\end{equation}
where
\begin{equation}
\alpha_B=
\left\{
\begin{array}{ll}
\tilde\alpha & (\mbox{ $s$-wave pairing})\\ 
\tilde\alpha+\frac{4}{3\pi}\sin (\pi\tilde\alpha )+
\frac{1}{6\pi}\sin (2\pi\tilde\alpha ) & (\mbox{ $d$-wave pairing})
\end{array}.
\right.
\label{alb}
\end{equation}
These expressions allow direct numerical calculations of the coefficients
$K_A,K_C,K_B$. In Fig.8, for example, we present the calculated dependence of
$K_C$ on the width of the pseudogap $W$ and correlation length of AFM
fluctuations (parameter $\kappa=v_F/xi^{-1}$). The appropriate dependences of
$K_A$ and $K_B$ are qualitatively quite similar. In particular, for
$\kappa=0$ we have just $K_B=K_C$ \cite{PS}.

\section{Physical Properties of Superconductors with Pseudogap.}

Ginzburg -- Landau expansion defines two characteristic lengths of
superconductors: the coherence length and penetration depth of magnetic 
field.  

The coherence length for given temperature $\xi(T)$ determines the 
characteristic scale of inhomogeneities of the order parameter $\Delta$:
\begin{equation}
\xi^2(T)=-\frac{C}{A}.
\label{xii}
\end{equation}
In the absence of the pseudogap:
\begin{eqnarray}
\xi_{BCS}^2(T)=-\frac{C_{0}}{A_{0}}, \\
\xi_{BCS}(T)\approx 0.74\frac{\xi_{0}}{\sqrt{1-T/T_{c}}},
\label{xi}
\end{eqnarray}
where $\xi_{0}=0.18v_{F}/T_{c}$.\ In our model:
\begin{equation}
\frac{\xi^2(T)}{\xi_{BCS}^2(T)}=\frac{K_{C}}{K_{A}}.
\label{xiii}
\end{equation}
The dependences of $\xi^2(T)/\xi_{BCS}^2(T)$ on the width of the pseudogap
$W$ and correlation length of fluctuations (parameter $\kappa$) for the
case $d$-wave pairing are shown in Fig.9. Note that the changes of coherence
length are relatively small.

For the penetration depth of a superconductor without the pseudogap we have:
\begin{equation}
\lambda_{BCS}(T)=\frac{1}{\sqrt{2}}\frac{\lambda_{0}}{\sqrt{1-T/T_{c}}},
\label{lamb}
\end{equation}
where $\lambda_{0}^2=\frac{mc^2}{4\pi ne^2}$ is penetration depth at $T=0$.\ 
In general case:
\begin{equation}
\lambda^2(T)=-\frac{c^2}{32\pi e^2}\frac{B}{AC}.
\label{lam}
\end{equation}
Then for our model:
\begin{equation}
\frac{\lambda(T)}{\lambda_{BCS}(T)}=
\left(\frac{K_{B}}{K_{A}K_{C}}\right)^{1/2}.
\label{lm}
\end{equation}
Graphical dependences of penetration depth for the case of $d$-wave pairing
are shown in Fig.10. 

Near $T_{c}$ the upper critical magnetic field $H_{c2}$ is defined via
Ginzburg -- Landau coefficients as: 
\begin{equation} 
H_{c2}=\frac{\phi_0}{2\pi\xi^2(T)}=-\frac{\phi_{0}}{2\pi}\frac{A}{C} ,
\label{Hc2} 
\end{equation} 
where $\phi_{0}=c\pi/e$ is magnetic flux quantum. 
Then the slope of the upper critical field close to $T_{c}$ is defined as:  
\begin{equation} 
\left|\frac{dH_{c2}}{dT}\right|_{T_c}=
\frac{24\pi\phi_{0}}{7\zeta(3)v_F^2}T_{c}
\frac{K_A}{K_C}. 
\label{dHc2}
\end{equation}
Graphic dependences of the slope of the upper critical field
$\left|\frac{dH_{c2}}{dT}\right|_{T_c}$, normalized to the slope at
$T_{c0}$, on the effective width of the pseudogap $W$ and parameter of
correlation length $\kappa$ for the case of $d$-wave pairing are shown in
Fig.11. It is seen that the slope for large enough correlation lengths
rapidly drops with the width of the pseudogap. However, for short enough
correlation lengths we can observe even some weak growth of this parameter
for small values of the pseudogap width. For fixed pseudogap width the
slope of $H_{c2}$ significantly grows as correlation length becomes smaller.

Finally, let us consider the discontinuity of specific heat at the transition
point:
\begin{equation} 
\frac{C_s-C_n}{\Omega}=\frac{T_c}{B}\left(\frac{A}{T-T_c}\right)^2,
\label{Cs}
\end{equation}
where $C_s,\>C_n$ are specific heats of superconducting and normal states,
$\Omega$ -- sample volume. For $T=T_{c0}$ (in the absence of pseudogap, 
$W=0$):  
\begin{equation} 
\left(\frac{C_s-C_n}{\Omega}\right)_{T_{c0}}=N(0)\frac{8\pi^2T_{c0}}{7\zeta(3)}.
\label{CsCn}
\end{equation}
Then the normalized discontinuity of specific heat in our model can be
expressed as:
\begin{equation}
\frac{(C_s-C_n)_{T_c}}{(C_s-C_n)_{T_{c0}}}=
\frac{T_c}{T_{c0}}\frac{K_A^2}{K_B}.
\label{cscn}
\end{equation}
Appropriate dependences on effective width of the pseudogap  $W$ and
parameter of correlation length $\kappa$ for the case of $d$-wave pairing are
shown in Fig.12. It is seen that specific heat discontinuity rapidly drops
with the growth of the pseudogap width and grows as correlation length of
AFM fluctuations becomes smaller.

For $s$-wave superconductor the dependences of physical properties are
quite similar, the only change is in larger scale of $W$ for which the
appropriate changes appear, corresponding to larger stability of isotropic
superconductors to partial ``dielectrization'' of electronic spectrum due
to pseudogap formation on ``hot patches'' of the Fermi surface \cite{PS,KSad}.

Among the physical characteristics, analyzed above, relatively detailed
experimental data are available for specific heat discontinuity \cite{Lor}.
In complete agreement with our conclusions, specific heat discontinuity for
$Bi-2212$ rapidly drops as the system moves to the underdoped region, where
the width of the pseudogap grows. According to Ref. \cite{Lor} the width of
the pseudogap (our parameter $2W$) changes from the values of the order of
$700K$ for hole concentration $p=0.05$ to the values of the order of
$T_c\sim 100K$ near optimal concentration $p=0.16$ and drops to zero for
$p=0.19$. A relation between the drop of specific heat discontinuity
and the growth of the pseudogap width is clearly observed. Unfortunately,
we do not know detailed enough data on concentration dependence of
correlation length of fluctuations and appropriate dependences of physical
characteristics of superconductors in pseudogap state. Qualitatively it is
obvious that correlation length grows as system moves to the underdoped 
region, so that the drop of specific heat discontinuity is natural also from
this point of view.

\section{Conclusion.}

In this paper we continued the study of anomalies of superconducting state
in the framework of rather crude model of the pseudogap state of two --
dimensional electronic system \cite{PS,KSad}, which is, however, in
qualitative agreement with a number of observed anomalies of electronic
structure of underdoped HTSC -- cuprates. In Refs. \cite{PS,KSad} we
considered rather unrealistic limit of infinite correlation length of
AFM short -- range order fluctuations, which allowed us to obtain an exact
analytic solution. Here we generalized our model to realistic case of finite
correlation lengths, taking into account, as in Refs. \cite{PS,KSad}, 
all diagrams of perturbation theory on electron interaction with
fluctuations of short -- range order. Our analysis was performed in a
standard (mean -- field in terms of Ref. \cite{KSad}) approach, assuming
self -- averaging property of superconducting order -- parameter over
fluctuations of the random field of AFM fluctuations. In Ref. \cite{KSad} 
we have shown that this assumption is not justified in the limit of
$\xi\to\infty$. At the same time this assumption is apparently well justified
for the case of $\xi\ll\xi_0$ (where $\xi_0$ is the coherence length of a
superconductor at $T=0$, i.e. the size of Cooper pairs). Thus, we are left
with rather complicated task of the accounting of non self -- averaging
effects for $\xi>\xi_0$. We have already mentioned that in real HTSC systems
in most cases $\xi\sim\xi_0$, so that effects of non self -- averaging 
superconducting gap, similar to those considered in Ref. \cite{KSad}, may be
very important, leading e.g. to characteristic ``tails'' in the temperature
dependence of the average gap for $T>T_c$ and the physical picture of
superconducting ``drops'' of Ref. \cite{KSad}.

Another serious simplification of our model is the assumption of static
(and Gaussian) nature of short -- range order of fluctuations. This
assumption may be justified only for high enough temperatures 
$T\gg\omega_{sf}$ (where $\omega_{sf}$ -- is characteristic scale of
spin fluctuations) \cite{Sch,KS}. Thus, the use of static approximation in
superconducting state for $T<T_c$ is rather doubtful. However, we think that
our simplified treatment allows us to describe most important effects of the
changes of the electronic spectrum (due to pseudogap formation on ``hot
patches'' of the Fermi surface) upon superconductivity. The account of spin
dynamics inevitably requires to drop the simple phenomenology of BCS model
and consider the microscopic nature of pairing interaction. It is doubtful
that such a program can be realized in near future. In particular, the 
problem of summation of all perturbation theory diagrams for the interaction
with dynamical spin fluctuations seems absolutely hopeless.

This work was partially supported by grants 99-02-16285 of the Russian
Foundation of Basic Research and REC-005 of CRDF, 
as well as by the projects 108-11(00-$\Pi$)
of the State Program ``Statistical Physics'' and 96-051 of the State Program
on HTSC. 

\newpage

\appendix

\section{Coordinate representation. Normal and anomalous Green's functions.}

Let us consider some technical details of derivation of recursion relations
for Gorkov equations (\ref{Gork}) -- (\ref{renpar}). It is sufficient to
limit the analysis to consideration of two flat parts of the Fermi surface,
orthogonal to $p_x$ -- axis, which are connected by the scattering vector 
${\bf Q}=(\pm 2p_F,0)$. Then the problem becomes purely one -- dimensional
as the velocity projection $v_y=0$ and the linearized electronic spectrum
$\xi_{p_x\mp p_F}=\pm v_Fp_x$ does not depend on $y$-component of electronic
momentum. For brevity in the following we just put $v_F=1$.  

Calculations simplify in coordinate representation \cite{Tch}, 
considering the electron propagation if the field of Gaussian AFM 
fluctuations $W(x)\neq W^*(x)$ (incommensurate case) with correlator:
\begin{equation}
<W^*(x)W(x')>=W^2e^{-\kappa|x-x'|}
\label{corW}
\end{equation}
Then electron propagators corresponding to normal and anomalous Green's
functions of a superconductor (\ref{GoFo}), take the form:
\begin{eqnarray}
G_{00}(x)=\int_{-\infty}^{\infty}\frac{dp_x}{2\pi}e^{ip_xx}G_{00}(p_x)=
-\frac{i}{2}\left(\frac{\varepsilon_n}{\sqrt{\varepsilon_n^2+|\Delta|^2}}
+\sigma_3sign(x)\right)e^{-\sqrt{\varepsilon_n^2+|\Delta|^2}|x|}\nonumber\\
F_{00}(x)=\int_{-\infty}^{\infty}\frac{dp_x}{2\pi}e^{ip_xx}F^+_{00}(p_x)=
\frac{\Delta^*}{\sqrt{\varepsilon_n^2+|\Delta|^2}}
e^{-\sqrt{\varepsilon_n^2+|\Delta|^2}|x|}
\label{GxFx}
\end{eqnarray}
where $\sigma_3=1$ for right moving particles, and $\sigma_3=-1$ for left
moving particles. Scattering by fluctuations transforms ``right'' particles
into ``left'' and vice versa. From (\ref{GxFx}) it is seen that the particle
moving along the path of the length $l$ produces the factor 
$e^{-\sqrt{\varepsilon_n^2+|\Delta|^2}l}$.

During calculation of specific diagrams it is convenient \cite{Tch} 
to change integration variables from coordinates of interaction vertices
$x_k$ to lengths of the paths $l_k$ traversed by the particle between separate
scatterings, fixing the total displacement $x-x'$. With interaction line 
connecting the vertices  $m$ and $n$ on electronic line we have to 
associate the factor:
\begin{equation}
W^2exp{(-\kappa|x_{m}-x_{n}|)}=W^2exp{(-\kappa|\sum_{k=m}
^{n-1}(-1)^kl_k|)}
\label{intline}
\end{equation}
Integration over all $l_k$ is performed from $0$ to $\infty$.

It is seen that the finite correlation length of fluctuations in a given
diagram leads to some ``damping'' of the appropriate transition amplitude
with distance traversed by an electron. The exact treatment of this effect
is difficult, but in Ref. \cite{KS} we used an obvious inequality: 
\begin{equation}
exp\Bigl(-\kappa|\sum_{k=m}^{n-1}(-1)^kl_k|\Bigr)>
exp\Bigl(-\kappa\sum_{k=m}^{n-1}l_k\Bigr)
\label{ineq}
\end{equation}
and replaced the exponential of (\ref{intline}) by the exponential from the
r.h.s. of (\ref{ineq}). This is equivalent to the replacement of correlator
of random fields (\ref{corW}) by similar expression, where in the exponent
we just replace the distance $|x-x'|$ by the total length of the path
traversed by the particle between scattering acts at $x$ and $x'$.
According to (\ref{ineq}) this procedure somehow overestimates the effect
of damping $\kappa$ in each diagram of perturbation theory. After such a
replacement the diagrams of all orders are easily calculated and precisely
reproduce the {\em Ansatz} of (\ref{Ansatz}) for the normal phase \cite{Tch}.
We have already mentioned that results obtained in this way, e.g. for the
density of states, are in good quantitative agreement with exact numerical
simulation of the one -- dimensional problem \cite{Kop,Mill}, which gives
additional support for our approximation strengthening qualitative estimates
of Ref. \cite{KS}. 

Let us use the same approximation during the analysis of diagrams of
perturbation theory in superconducting phase, which are built upon
propagators (\ref{GxFx}). In this case the role of interaction reduces just
to the appearance of additional factor of $e^{-\kappa l_k}$ in each
normal or anomalous Green's function (\ref{GxFx}), surrounded by the given
interaction line, or (which is the same) to the addition of $\kappa$ to 
$\sqrt{\varepsilon_n+|\Delta|^2}$ in the exponential of each Green's 
function. Making transformation back to the momentum representation it is
easily seen that the contribution of an arbitrary diagram of the higher
order of perturbation theory is determined by the product of the
appropriate number of normal and anomalous Green's functions of the form: 
\begin{equation} 
G_{0k}(p)=-\frac{i\varepsilon_n\frac{\varepsilon_k}
{\sqrt{\varepsilon_n+|\Delta|^2}}+(-1)^k\xi_p}{\varepsilon_k^2+\xi_p^2};
\quad
F^+_{0k}(p)=\frac{\Delta^*\frac{\varepsilon_k}
{\sqrt{\varepsilon_n+|\Delta|^2}}}{\varepsilon_k^2+\xi_p^2};
\label{GokFok}
\end{equation}
where $\varepsilon_k=\sqrt{\varepsilon_n+|\Delta|^2}+k\kappa$, while $k$ -- 
is the number of interaction lines, surrounding the given Green's function.
The factor of $(-1)^k$ is due to scattering transforming ``right'' particles
into  ``left'' and vice versa. Introducing the renormalized frequency and gap
as in (\ref{renpar}), we can see that (\ref{GokFok}) reduces to the standard
form (\ref{GkFk}), which completes the justification of our recursion
procedure (\ref{Gork}), (\ref{renpar}).

\newpage
\begin{center}
{\bf Figure Captions:}
\end{center}

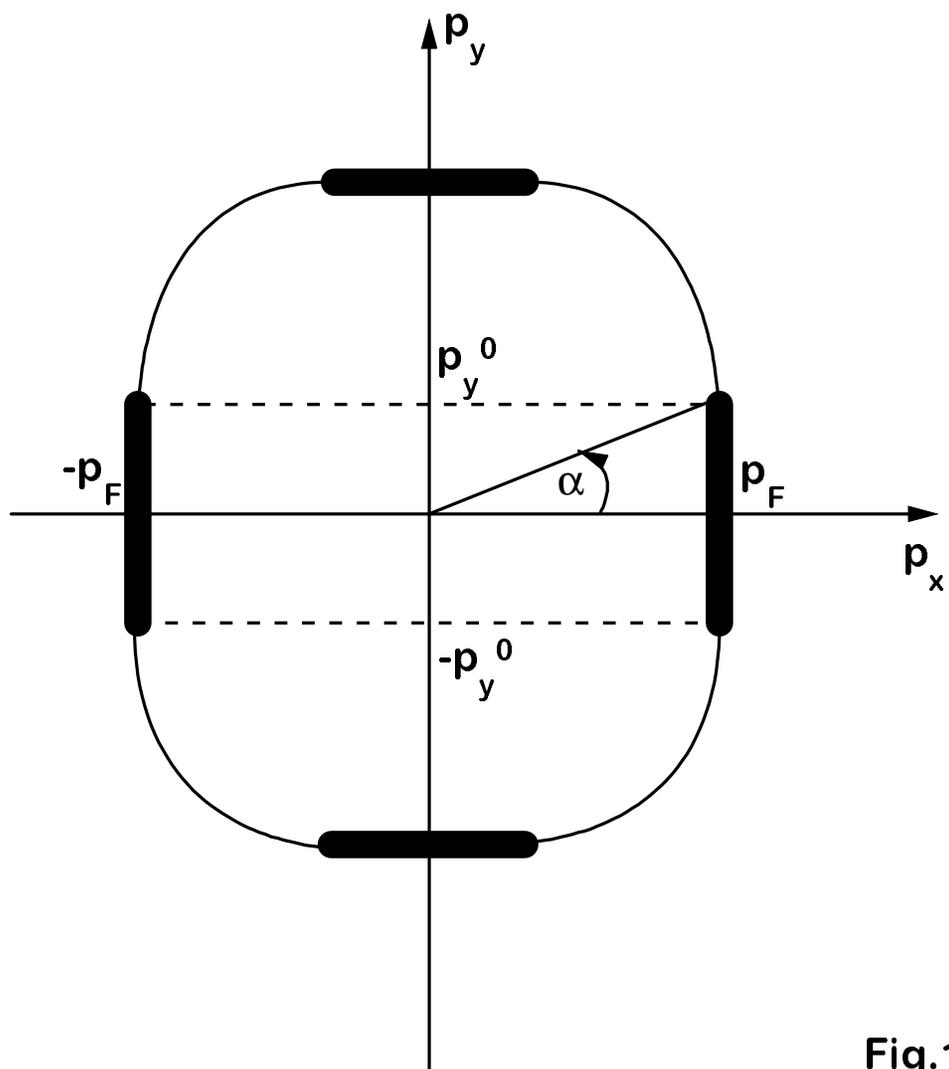
Fig.1. Fermi surface of two -- dimensional system.
``Hot patches'' are shown by thick lines of the width $\sim \xi^{-1}$. 

Fig.2. Diagrammatic representation of recursion relation for one -- particle
Green's function.

Fig.3. Diagrammatic representation of recursion relations for Gorkov
equations.

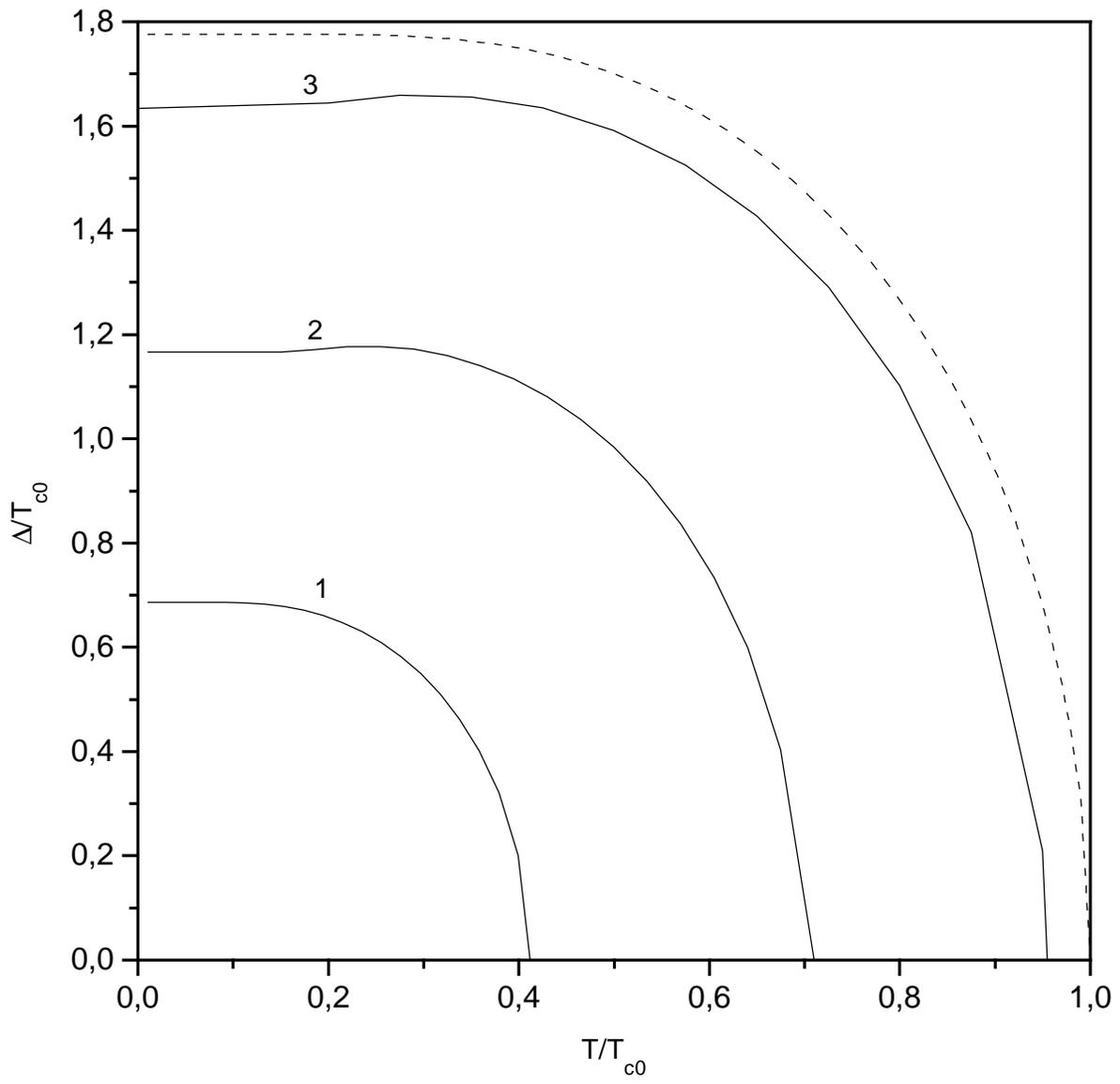
Fig.4. Temperature dependence of superconducting energy gap for the case of
$s$-wave pairing and different values of correlation length
(parameter $\kappa=v_F\xi^{-1}$) of AFM fluctuations, calculated for
$\lambda=0.4,\ \frac{\omega_c}{W}=3$:

$\frac{\kappa}{W}$=0\ (1);\quad 1.0\ (2);\quad 10.0\ (3).

Dashed line --- $\Delta(T)$ in the absence of the pseudogap.

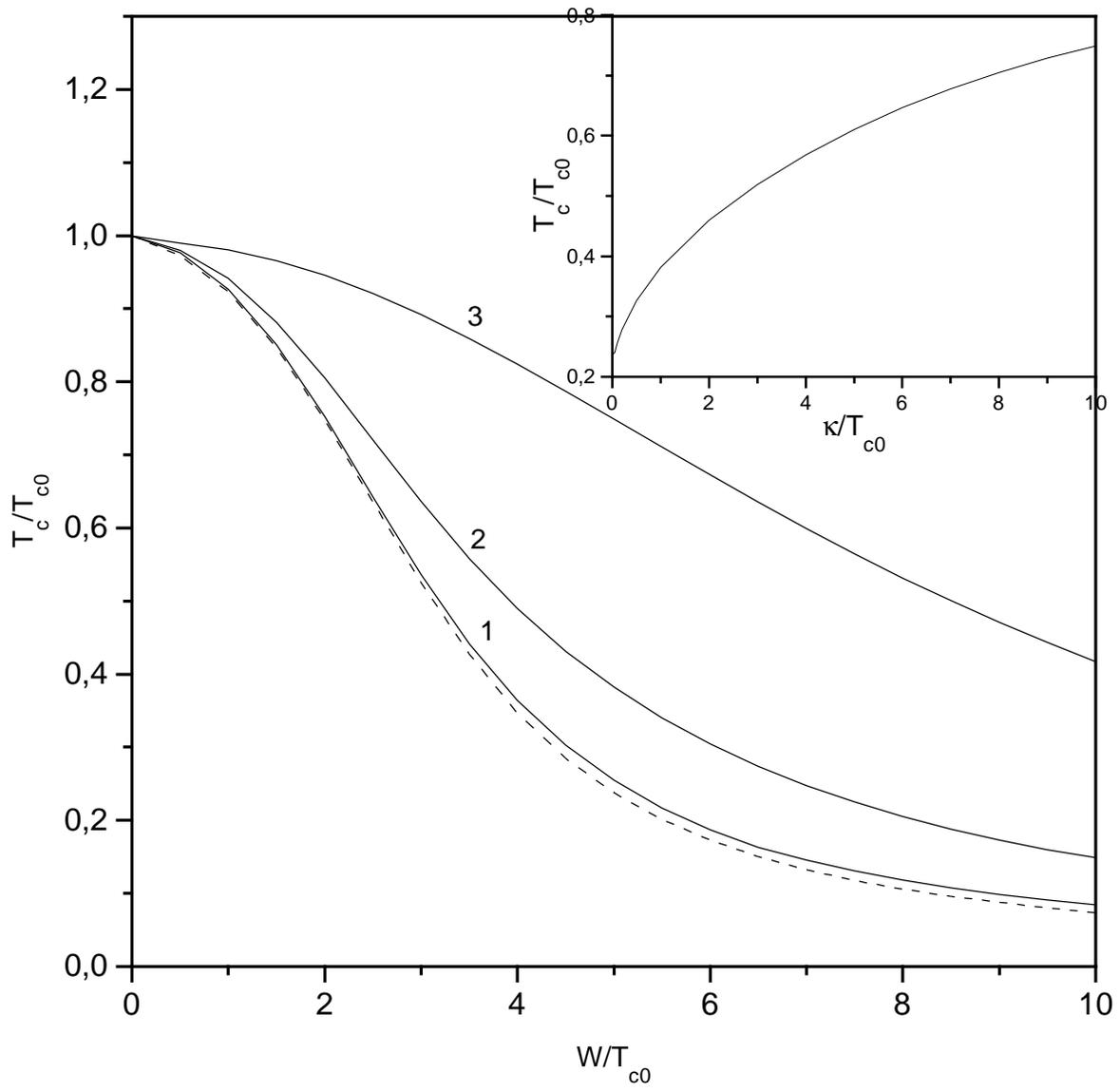
Fig.5. Dependence of superconducting transition temperature on the width of
the pseudogap $W$ and correlation length of AFM fluctuations
(parameter $\kappa=v_F\xi^{-1}$):

$\frac{\kappa}{W}$=0.1\ (1);\quad 1.0\ (2);\quad 10.0\ (3).

Dashed line --- $\kappa=0$ \cite{PS}.

At the insert: dependence of $T_c$ on $\kappa$ for $\frac{W}{T_{c0}}=5$.

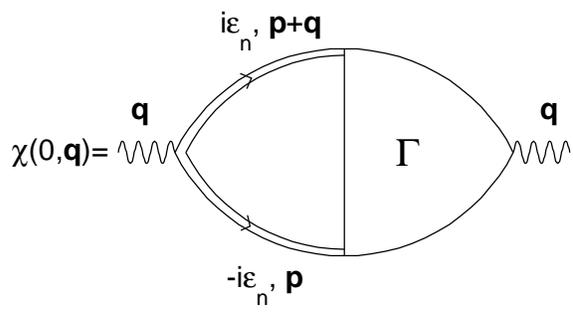
Fig.6. Diagram for the generalized Cooper susceptibility.

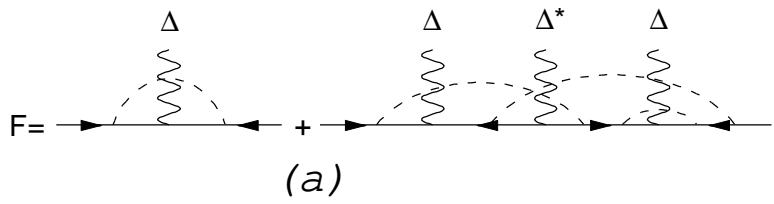
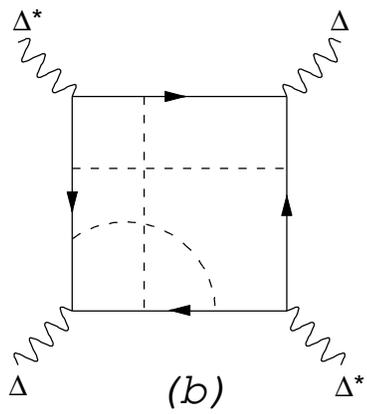
Fig.7. (a) -- Diagram series for anomalous Green's function, dashed lines 
-- AFM fluctuations. (b) -- Diagram determining $K_B$.

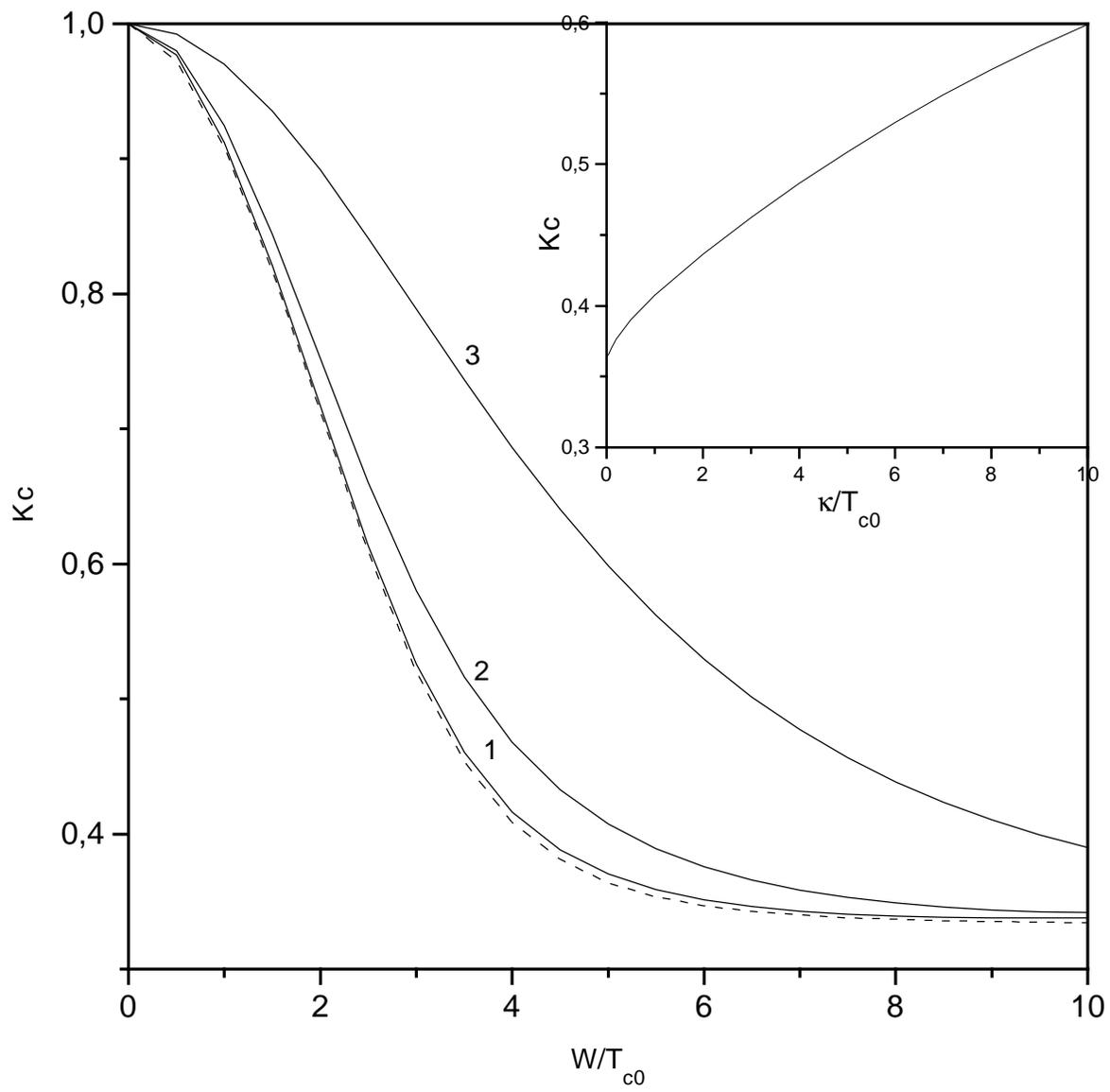
Fig.8. Dependence of $K_C$ on the width of the pseudogap
$W$ and correlation length of AFM  fluctuations
(parameter $\kappa=v_F\xi^{-1}$):

$\frac{\kappa}{W}$=0.1\ (1);\quad 1.0\ (2);\quad 10.0\ (3).

Dashed line --- $\kappa=0$ \cite{PS}.

At the insert: dependence of $K_C$ on $\kappa$ for $\frac{W}{T_{c0}}=5$.

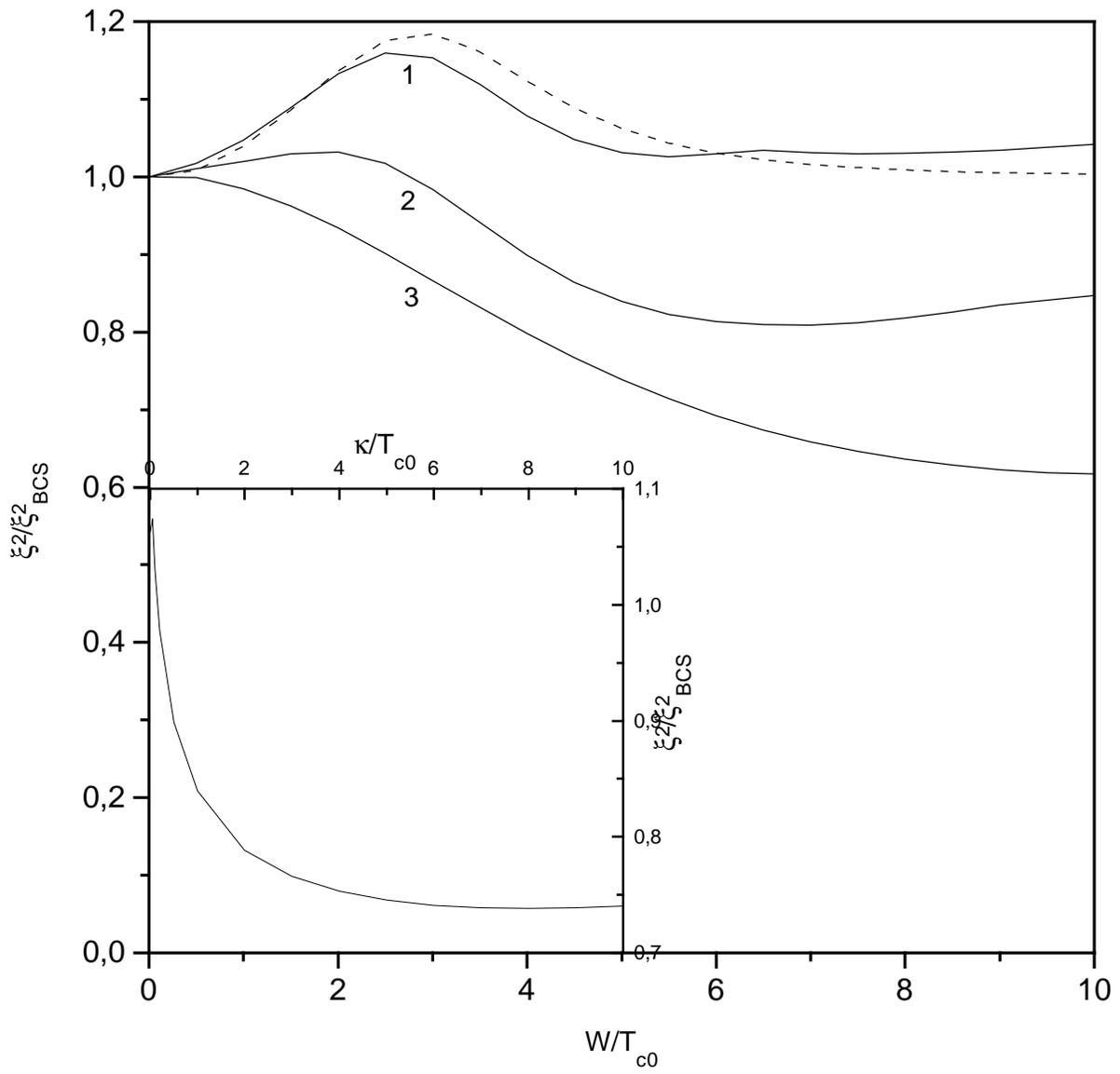
Fig.9. Dependence of supercoducting coherence length on the width of the 
pseudogap $W$ and correlation length of AFM fluctuations (parameter 
$\kappa=v_F\xi^{-1}$):

$\frac{\kappa}{W}$=0.1\ (1);\quad 1.0\ (2);\quad 10.0\ (3).

Dashed line --- $\kappa=0$ \cite{PS}.

At the insert: dependence of coherence length on $\kappa$ for 
$\frac{W}{T_{c0}}=5$.

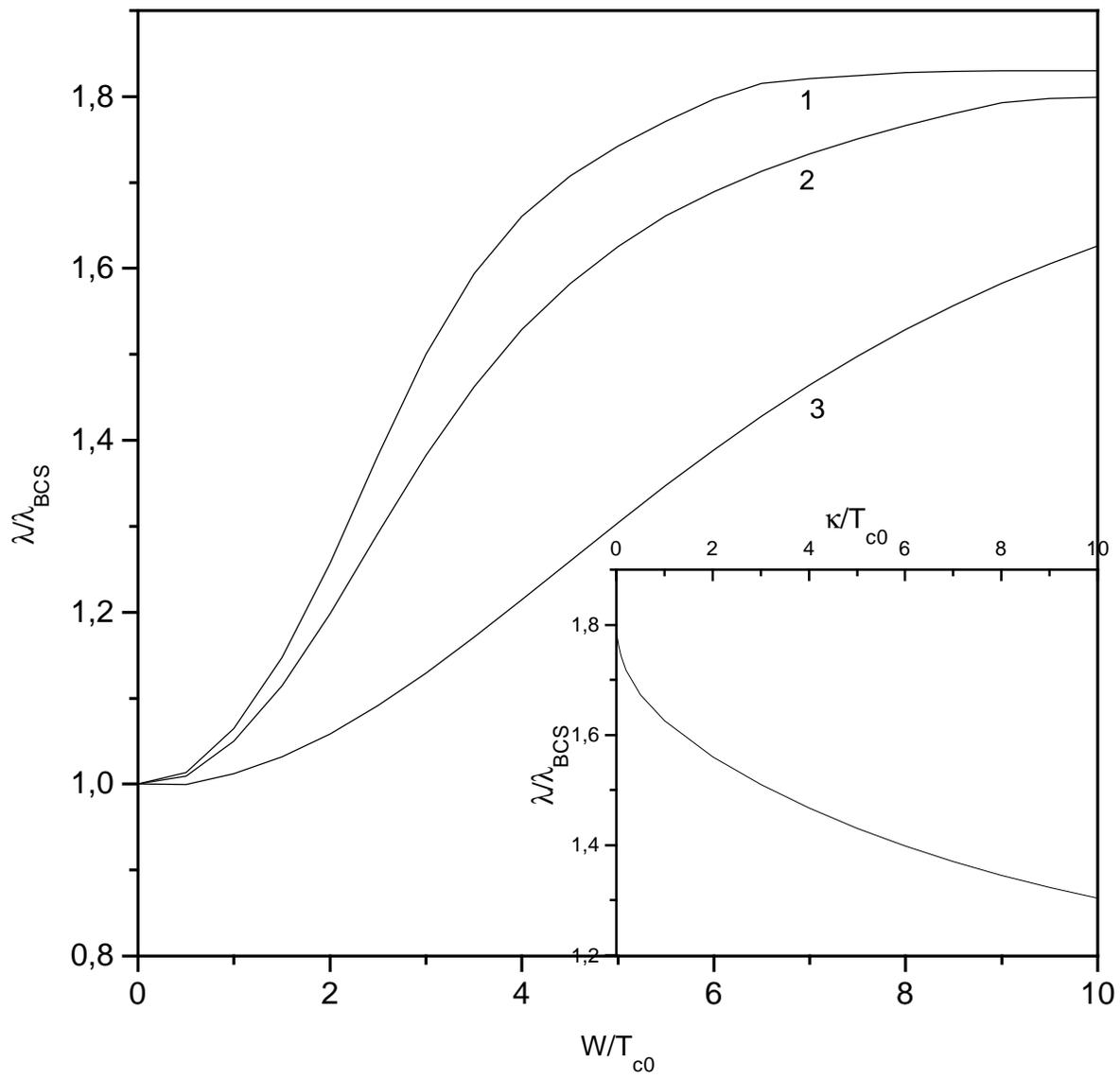
Fig.10. Dependence of penetration depth on the width of the pseudogap
$W$ and correlation length of AFM  fluctuations
(parameter $\kappa=v_F\xi^{-1}$):

$\frac{\kappa}{W}$=0.1\ (1);\quad 1.0\ (2);\quad 10.0\ (3).


At the insert: dependence of penetration length on $\kappa$ for 
$\frac{W}{T_{c0}}=5$.

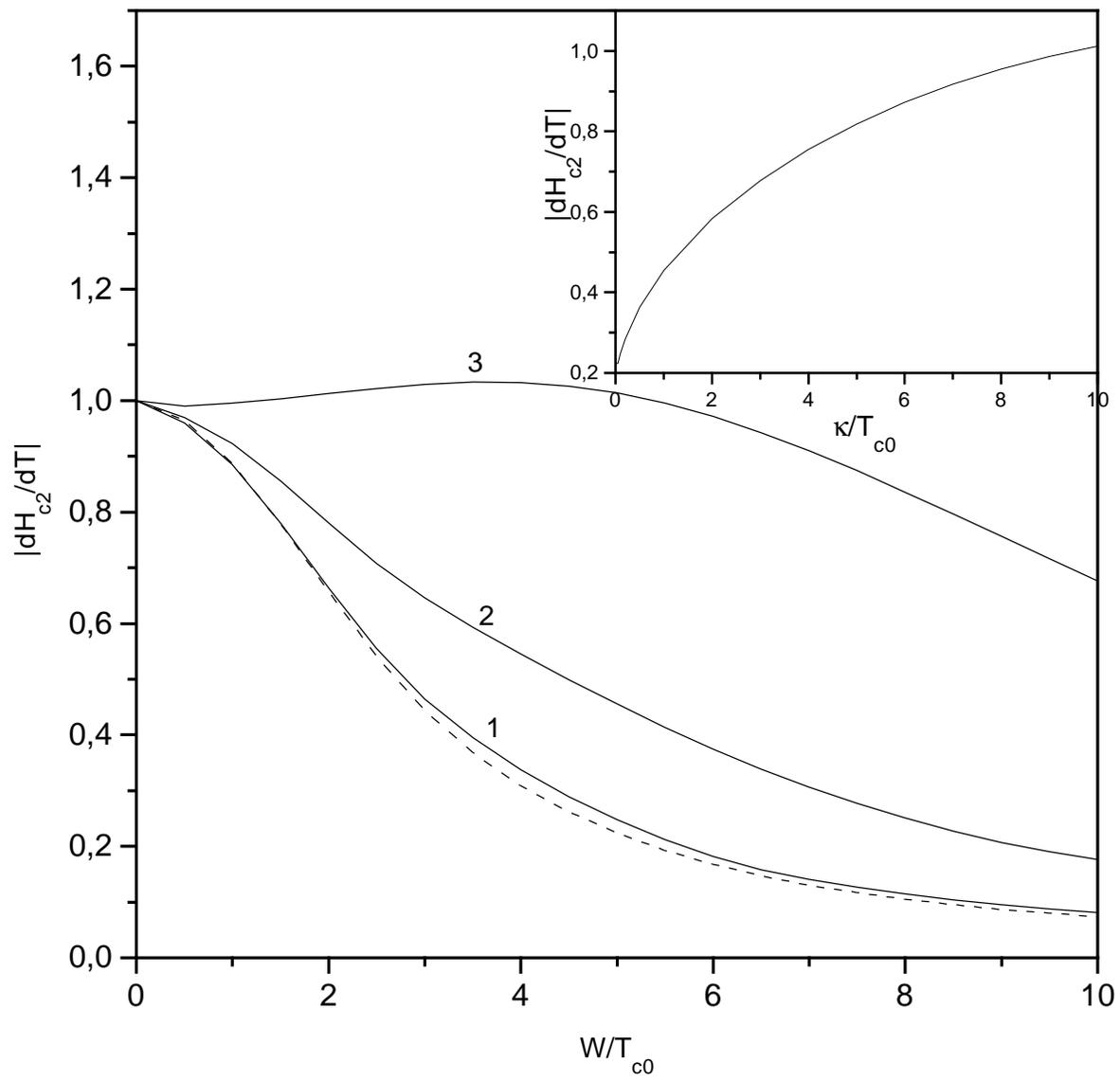
Fig.11. Dependence of the slope of the upper critical field on the width of
the pseudogap $W$ and correlation length of AFM fluctuations (parameter 
$\kappa=v_F\xi^{-1}$):

$\frac{\kappa}{W}$=0.1\ (1);\quad 1.0\ (2);\quad 10.0\ (3).

Dashed line --- $\kappa=0$ \cite{PS}.

At the insert: dependence of the slope of $H_{c2}$ on $\kappa$ for 
$\frac{W}{T_{c0}}=5$.

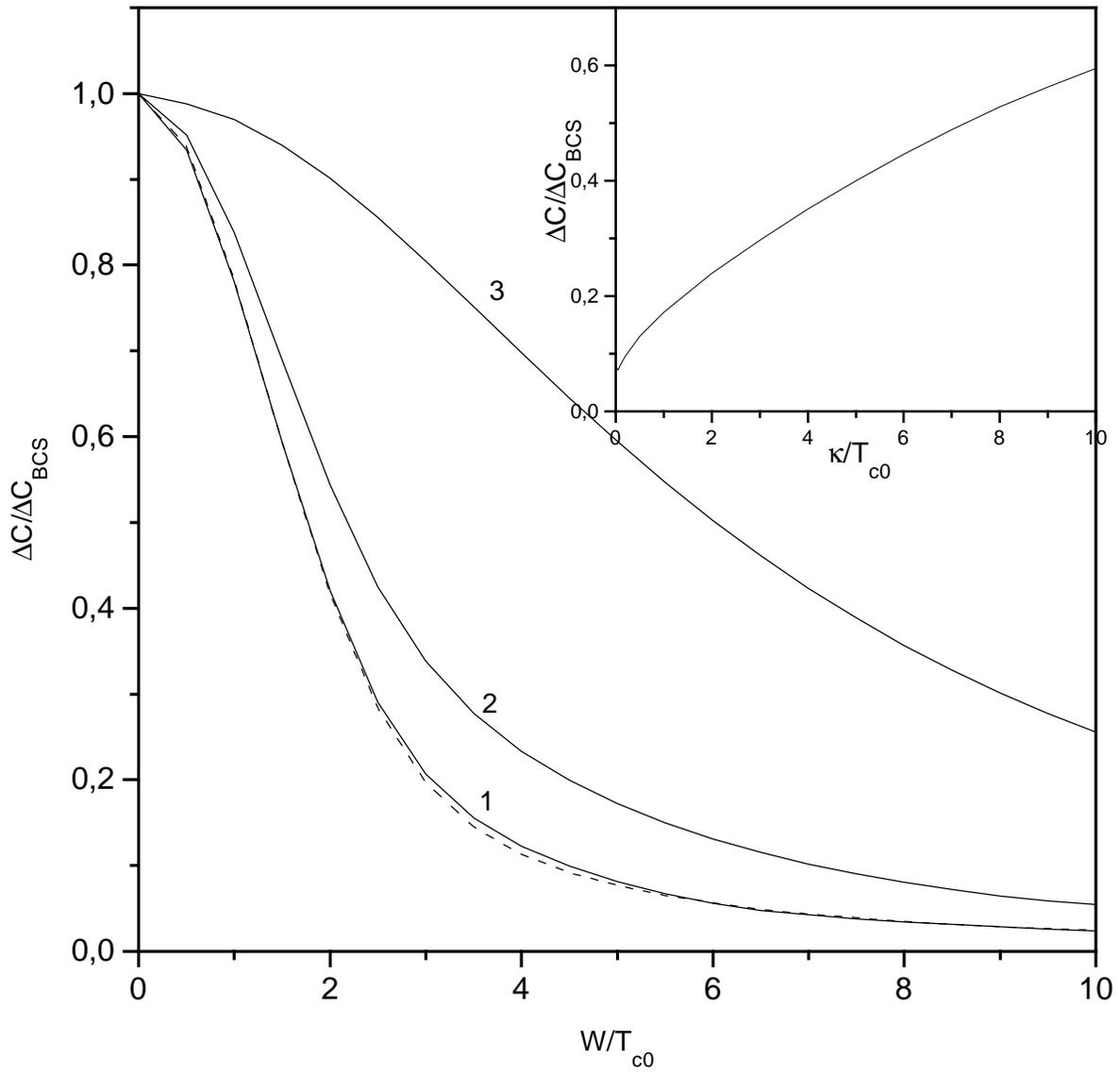
Fig.12. Dependence of specific heat discontinuity on the width of the
pseudogap $W$ and correlation length of AFM fluctuations 
(parameter $\kappa=v_F\xi^{-1}$):

$\frac{\kappa}{W}$=0.1\ (1);\quad 1.0\ (2);\quad 10.0\ (3).

Dashed line --- $\kappa=0$ \cite{PS}.

At the insert: dependence of specific heat discontinuity on $\kappa$ 
for $\frac{W}{T_{c0}}=5$.

\end{document}